\documentclass[iop]{emulateapj}

\usepackage{natbib}
\shorttitle{Fast Modes and Dusty Horseshoes}
\shortauthors{Mittal and Chiang}

\begin{document}

\title{FAST MODES AND DUSTY HORSESHOES IN TRANSITIONAL DISKS}

\author{Tushar Mittal\altaffilmark{1} and Eugene Chiang\altaffilmark{1,2}}
\altaffiltext{1}{Department of Earth and Planetary Science, 307 McCone Hall, University of California, Berkeley, CA 94720-4767}
\altaffiltext{2}{Department of Astronomy, 501 Campbell Hall, University of California, Berkeley, CA 94720-3411}

\begin{abstract}

The brightest transitional protoplanetary disks are often azimuthally
asymmetric: their mm-wave thermal emission peaks strongly on one side.
Dust overdensities can exceed $\sim$100:1, while gas
densities vary by factors less than a few.
We propose that these remarkable ALMA
observations---which may bear on how
planetesimals form---reflect a gravitational global mode in the gas disk. 
The mode is (1) fast---its pattern speed equals the disk's mean Keplerian
frequency; (2) of azimuthal wavenumber $m=1$, displacing the host star from
the barycenter; and (3) Toomre-stable. We solve for gas streamlines 
including the indirect stellar potential in the
frame rotating with the pattern speed, under the drastic simplification that
gas does not feel its own gravity.
Near co-rotation, the gas disk takes the form of a horseshoe-shaped annulus. 
Dust particles with aerodynamic stopping times much shorter or much longer
than the orbital period are dragged by gas toward the 
horseshoe center.
For intermediate stopping times, dust converges toward a $\sim$45$^\circ$-wide
arc on the co-rotation circle. 
Particles that
do not reach their final accumulation points within disk
lifetimes, either because of gas turbulence or long particle drift times,
conform to horseshoe-shaped gas streamlines.  
Our mode is not self-consistent because we neglect gas self-gravity;
still, we expect that trends between accumulation location and particle size,
similar to those we have found,
are generically predicted by fast modes 
and are potentially observable.
Unlike vortices, global modes are not restricted in
radial width to the pressure scale height; their large radial
and azimuthal extents may better match observations.

\end{abstract}
\keywords{protoplanetary disks --- accretion, accretion disks --- hydrodynamics --- celestial mechanics}

\section{INTRODUCTION}
Transitional protoplanetary disks possess inner cavities
nearly devoid of dust (e.g., \citealt{2014arXiv1402.7103E}).
Outside these cavities, in some of the brightest disks,
radio images reveal that dust is not axisymmetric,
but clumps strongly to one side.
At dust continuum wavelengths, surface brightness contrasts range from
values on the order of a few \citep{2009ApJ...704..496B, 2012A&A...547A..84T, 2013ApJ...775...30I, 2014ApJ...783L..13P} to $\sim$30 (HD 142527; \citealt{2013Natur.493..191C};
\citealt{2013PASJ...65L..14F}) to $\sim$130 (Oph IRS 48; \citealt{2013Sci...340.1199V};
\citealt{2014A&A...562A..26B}). These contrasts are lower bounds where
observations are unresolved. Unlike dust, overdensities in CO gas are
limited to factors of a few at most \citep{2013Sci...340.1199V,2014A&A...562A..26B,2014arXiv1410.8168P}.

The lopsided dust concentrations---which may be
giving us a first empirical glimpse into the process of
planetesimal formation---cry out to be understood. One proposed mechanism is
dust trapping by gas vortices (e.g., \citealt{2013ApJ...775...17L}; \citealt{2014arXiv1405.2790Z}).
The inner rims of transitional disks may be subject to the Rossby wave
instability which can spawn anticyclonic vortices 
(e.g., \citealt{2014FlDyR..46d1401L}). Whether
vortices are radially
and azimuthally large enough to match the imaging data is debatable.

We submit here a different explanation, one that follows from a simple 
fact: a lopsided disk moves the barycenter away from the host star.
The gravitational potential therefore includes an
``indirect'' term (e.g., \citealt{1999ssd..book.....M}). Our paper
solves for gas streamlines including the
indirect potential, in the limit that gas does not feel
its own gravity. We will discover 
in this limit
that gas 
occupies a radially wide, horseshoe-shaped annulus that readily concentrates
dust grains into patterns similar to those observed.
Concentration is by gas drag, not dust self-gravity.

Our proposal is that the gas disk exhibits a global ``fast'' mode of azimuthal wavenumber $m=1$, whose lopsidedness shifts
the barycenter off the star. ``Fast'' means
that the disk's pattern speed equals its mean Keplerian motion, not the apsidal
precession frequency that characterizes $m=1$ ``slow'' modes
(cf.~\citealt{2001AJ....121.1776T}).
Fast modes are more attractive than slow modes 
for shaping dust grain trajectories
and explaining the radio observations,
since the non-inertial forces arising from a slow mode's
pattern speed are too small to compete with the substantial gas drag forces
felt by dust.

We compute gas streamlines (test particle orbits) analytically in \S2 and numerically in \S3. How dust concentrates 
aerodynamically
is described in \S4.
Limitations of our calculations---notably
our neglect of gas
self-gravity, which prevents us from
computing fast modes self-consistently---are discussed in \S5.

\section{TEST PARTICLE DYNAMICS}

\begin{figure}
\epsscale{1.}
\figurenum{1}
\plotone{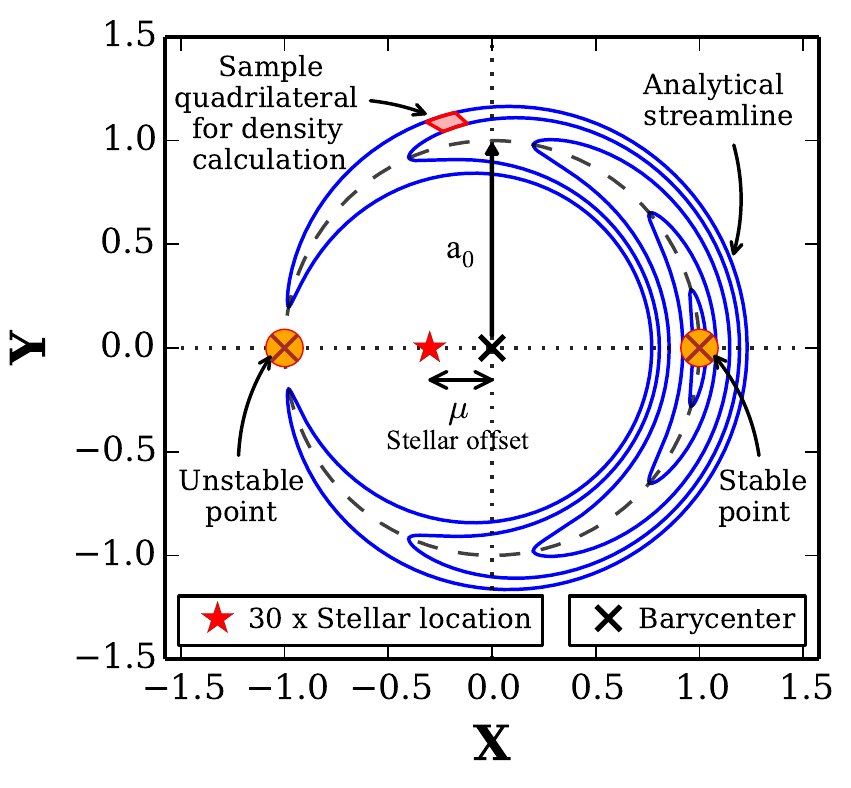}
\label{Fig1} 
\caption{In the offset stellar potential, closed and non-crossing orbits take the
form of horseshoes enclosing the stable point, and are plotted here using equation 
(\ref{B_eqn}) for $B$ = \{0.02, 0.12, 0.4, 0.7, 0.99\}. We show a sample quadrilateral used to construct the gas density.}
\end{figure}

We consider analytically the dynamics of a test particle in the gravitational potential of a star with a prescribed motion. 
The star has mass $M_\ast$ and executes a circular orbit of radius $\mu$ with fixed angular frequency $\Omega$ about the origin
(star-disk barycenter).
This orbit represents the star's ``reflex motion'' in response
to the $m=1$ component of the disk's potential.
We work in the frame rotating with the star and centered on the origin (Figure \ref{Fig1}). In our unit system, the gravitational constant
times the stellar mass $GM_\ast = 1-\mu$, $\Omega = 1$, and $\mu \ll 1$.
Our problem is identical to the standard restricted three-body problem,
except that the test particle does not feel the gravity
of the secondary (read: disk).
Because we assume the star's motion arises from the disk potential but neglect
the latter when computing the test particle's motion, our
calculations are not self-consistent; we proceed anyway in the hope that
some of the grosser, qualitative features of
our model will survive a more careful study.

In Cartesian coordinates, the test particle obeys
\begin{eqnarray}
\label{Eqn_motion_x}
\ddot{x} - 2\Omega\dot{y} & = & \frac{\partial U}{\partial x} \\
\label{Eqn_motion_y}
\ddot{y} + 2\Omega\dot{x} & = & \frac{\partial U}{\partial y} 
\end{eqnarray}
where $U$ is the celestial mechanician's centrifugal potential plus
the potential of the star (offset
from the origin by $\mu$ on the negative $x$-axis):
\begin{equation}
U = \frac{\Omega^2}{2} (x^2 + y^2) + \frac{1-\mu}{\sqrt{(x+\mu)^2 + y^2}} \,.
\end{equation}
Equilibrium points, where gravity
and centrifugal forces balance, lie along the $x$-axis.
Setting $\ddot{x} =\ddot{y} =\dot{x} = \dot{y} = y = 0$
in equations (\ref{Eqn_motion_x}) and (\ref{Eqn_motion_y}),
we find
\begin{equation}
\label{stable_pt1}
\frac{\partial U}{\partial x} = x_{\rm eq} - \frac{(x_{\rm eq}+\mu)(1-\mu)}{|x_{\rm eq}+\mu|^3} = 0 \,.
\end{equation}
To order $\mu$, the two equilibrium points are located at
\begin{equation}
\label{stable_pt_eqn}
x_{\rm eq} = 1 - \mu \,\,\, {\rm and} \,\, -1 - \mu/3 \,.
\end{equation}

Assuming that small displacements about these equilibria
evolve with time $t$ according to $\exp (\lambda t)$, we 
linearize and combine equations (\ref{Eqn_motion_x}) and (\ref{Eqn_motion_y})
to arrive at the relation for the eigenfrequency $\lambda$:
\begin{eqnarray}\label{max_mu1}
\lambda^4 & + & \left[ 2 - \frac{1-\mu}{(x_{\rm eq} + \mu)^3} \right]\lambda^2 \nonumber \\ & + & \left[ 1 + \frac{1-\mu}{(x_{\rm eq} + \mu)^3} -2 \frac{(1-\mu)^2}{(x_{\rm eq} + \mu)^6} \right] =0 \,.
\end{eqnarray}
The equilibrium point 
at $x_{\rm eq} < 0$
is unstable (${\rm Re} \, \lambda > 0$).
By contrast,
the equilibrium point at $x_{\rm eq} = 1 -\mu$ is stable
for $\mu < 1/10$ and is characterized by two frequencies of oscillation: 
one nearly equal to the mean motion $\Omega$ (and representing the usual
epicyclic motion of an eccentric orbit), and another
\begin{equation}\label{slow}
|\lambda| = \sqrt{3\mu} \, \Omega \,.
\end{equation}
This slower frequency describes 
the libration of the particle on trajectories
that are shaped like horseshoes. To sketch these
horseshoes, we switch to polar coordinates
where the equations of motion read
\begin{eqnarray}
\label{Eqn_motion_r}
\ddot{r} - r\dot{\theta}^2 -2\Omega r\dot{\theta} = \frac{\partial U}{\partial r} \\
\label{Eqn_motion_th}
r\ddot{\theta} + 2\dot{r}\dot{\theta} +2\Omega\dot{r} = \frac{1}{r}\frac{\partial U}{\partial \theta} 
\end{eqnarray}
with
\begin{eqnarray}
U & = &  \frac{\Omega^2}{2} r^2 + \frac{1-\mu}{\sqrt{r^2 + \mu^2 + 2 r \mu \cos \theta}} 
\\
\label{U_approx} & \simeq & 
\frac{3}{2} + \frac{3}{2}\Delta^2 - \mu(1 + \cos\theta) \,.
\end{eqnarray}
In equation (\ref{U_approx}), we have expanded $U$ in 
small radial (but not small azimuthal) displacements 
$\Delta \equiv r - 1$ about the stable point.
We insert (\ref{U_approx}) into 
(\ref{Eqn_motion_r}) and (\ref{Eqn_motion_th}),
keeping only leading-order terms and taking $d/dt \ll \Omega$ to
filter out fast epicycles; then
\begin{eqnarray}
\label{Eqn_motion_r_approx}
-2\dot{\theta}  =  3\Delta \\
\ddot{\theta}  =  -6\mu\frac{\partial}{\partial \theta} \sin^{2}(\theta/2) \,.
\label{Eqn_motion_th_approx}
\end{eqnarray}
Multiplying (\ref{Eqn_motion_th_approx}) by $\dot{\theta}$, integrating over 
time, and substituting (\ref{Eqn_motion_r_approx}),
we obtain the following ``shape function'':
\begin{equation}\label{B_eqn}
\Delta^2 = \frac{16}{3}\mu[B - \sin^{2}(\theta/2)] \,.
\end{equation}
Here $B$ is a constant of integration that takes values between 0 (zero amplitude 
libration at $\theta = 0$) and 1 (maximal libration). Figure \ref{Fig1} samples 
several $B$-values; evidently the shape function (\ref{B_eqn}) traces 
horseshoe-shaped trajectories enclosing the stable point. The horseshoes
are radially widest at $\theta = 0$:
\begin{equation}\label{Disk_width_scale}
\max \Delta = \sqrt{\frac{16B\mu}{3}} \,.
\end{equation}
This width can be a large fraction
of the radius; e.g., for $B=1$ and $\mu = 0.01$ (characteristic
of a disk whose mass is $\sim$$0.01\: M_\ast$),
the full width is $2\max\Delta = 0.46$.

Although the libration trajectories are shaped like horseshoes,
the square-root scalings in (\ref{slow})
and (\ref{Disk_width_scale}) for the libration frequency
and width characterize tadpoles in the standard three-body problem. This kinship between our horseshoe orbits
and conventional tadpoles
is expected. In both types of motion, the turning points
(located farthest from the
positive $x$-axis) are effected by the offset host star which exerts
torques on the test
particle that deflect it back toward the $x$-axis. 

\section{CONSTRUCTING A MODEL GAS DISK}
We construct a model gas disk based on ideas developed by 
\citet{1977ApJ...216..822P}
and applied in many galactic
contexts (e.g., \citealt{1991MNRAS.252..210B}; \citealt{2007ApJ...668..236C}). Gas is assumed to occupy closed, non-intersecting
test particle orbits.
Crossing orbits are forbidden because they lead
to shocks and energy dissipation. Gas streamlines approximate
test particle orbits insofar
as gas sound speeds are smaller than orbital velocities
and gas self-gravity is negligible.

Working in the rotating frame of \S2, we set $\mu = 0.01$,
$M_\ast = M_\odot$, and $\Omega = \sqrt{GM_{\sun}/a_0^3}$,
where $a_0 = 100$ AU is the disk's characteristic radius.
The star is positioned
at $x_\ast = -\mu a_0 = -1 \, {\rm AU}$ and $y_\ast=0$.
Starting on the $x$-axis, we launch a test particle
with an initial velocity $\dot{y} < 0$ ($\dot{x} = 0$)
and integrate its trajectory until it re-crosses the $x$-axis.  
The final position
and velocity are required to match initial values to within 1 part in $10^{4}$;
the initial $\dot{y}$ is varied until these conditions are met. 
The integrations 
\begin{figure*}[tp]
\epsscale{1}
\figurenum{2}
\plotone{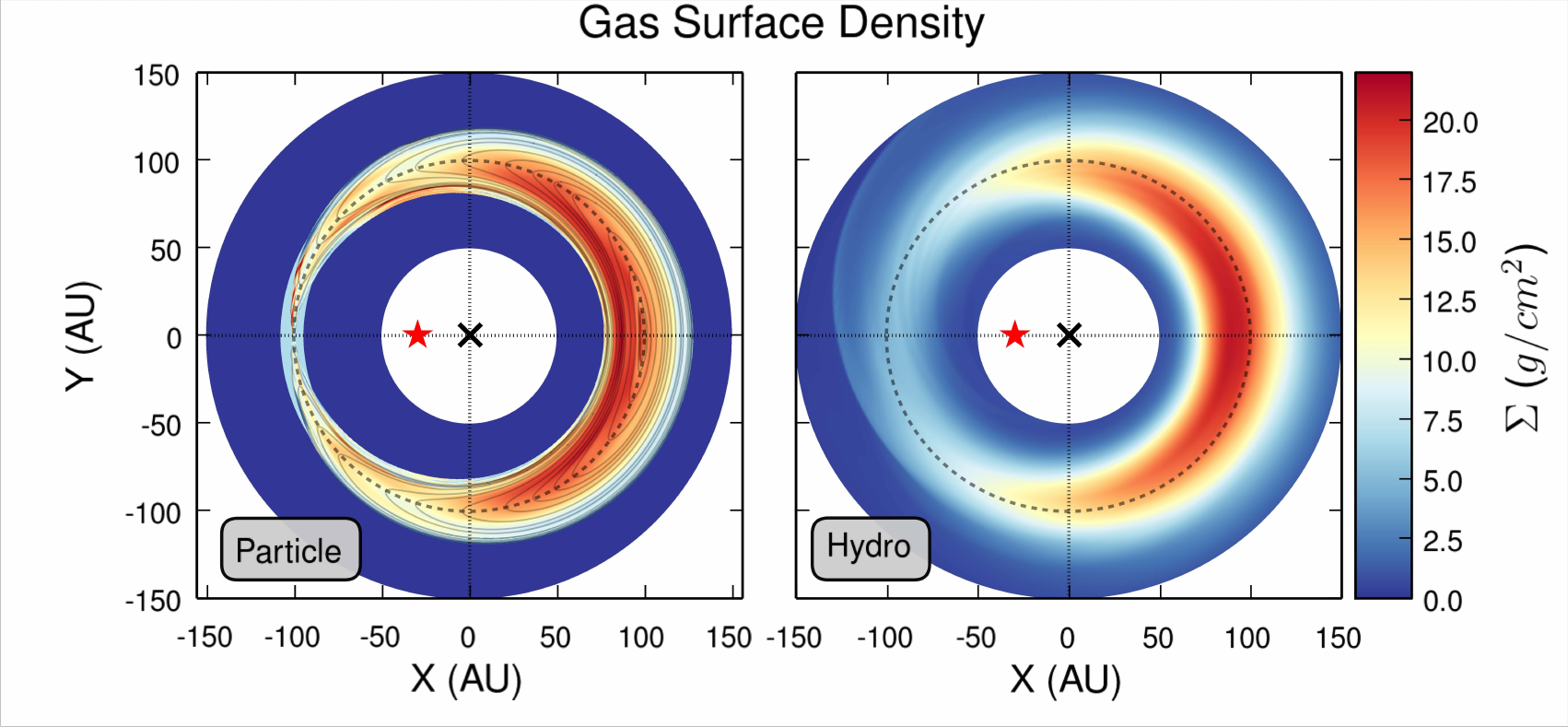}
\label{Fig2} 
\caption{{\it Left}: Gas surface density computed with our test-particle
method, for $\mu = 0.01$, $a_0 = 100$ AU (dashed
circle), and $M_{\rm disk} = 0.028 M_\odot$. Test particle orbits are overlaid
as contours.
The Toomre parameter $Q = c_{\rm s}\Omega / \pi G \Sigma \sim 2$--4
for a sound speed $c_{\rm s} \approx 0.37$ km/s (temperature 40 K).
Across the horseshoe-shaped annulus, $\Sigma$ varies by factors
$\lesssim 3$. {\it Right}: Gas surface density computed
with the hydrocode \texttt{PLUTO}, using parameters identical
to those in panel (a). We use a polar grid
that spans 0 to $2\pi$ in azimuth and 0.2$a_0$ to $2a_0$ in radius;
a grid resolution of 400 $\times$ 400 cells; outflow
boundary conditions; and an isothermal $c_{\rm s} = 0.37$ km/s.
Initially the disk is centered on the origin (frame rotation axis), 
extends from 0.9$a_0$ to 1.1$a_0$, and has uniform density.
The snapshot is taken after 80 orbits (evaluated at $a_0$), with
qualitatively similar results from 20--80 orbits; at earlier times,
strong transients afflict the disk, and at later times,
the disk has lost much of its mass through the outer boundary.
We are encouraged by the agreement between the test-particle
and hydrocode methods.}
\end{figure*}
\noindent are performed using Python's 
explicit Runge-Kutta integrator of order 8(5,3) (\texttt{dop853}) with an
accuracy setting of $10^{-8}$. A total of 22 initial $x$-positions,
distributed between the stable point at $x=x_{\rm eq}=99.3$ AU and $x=x_{\rm max}=127.4$ AU are selected for the construction of 22 orbits/streamlines; see Figure
\ref{Fig2}.
Their shapes match well the horseshoe orbits found analytically in
\S2. In particular, we confirm that equations (\ref{slow})
and (\ref{Disk_width_scale}) for the libration frequency
and radial width are obeyed.

Each streamline is defined by $N=300$ points marking the
test particle's position recorded at equal intervals of time.
To compute the gas surface density, we 
assign each streamline a mass
$\propto \cos [(\pi/2) (x-x_{\rm eq}) / (x_{\rm max}-x_{\rm eq})]$,
where $x$ identifies the streamline's starting position;
we assign more mass to inner horseshoes than
outer ones. Lines are drawn between adjacent points on neighboring
streamlines so that the entire space is tiled by quadrilaterals (see
Figure \ref{Fig1}). Each quadrilateral
of area $dA$ is assigned a surface density $\Sigma = dM/dA$,
where $dM =$ streamline mass / $N$.

For the most part this procedure yields surface densities
that vary smoothly from quadrilateral to quadrilateral.
However, we find a few spikes in density near the disk's edges
(outermost horseshoes). These are numerical artifacts that arise from our particular choice of streamlines. Because
our gravitational potential is only weakly non-axisymmetric,
it admits a large variety of orbits that satisfy our tolerance criteria for being closed and non-intersecting. 
Our selection of
streamlines is thus not unique.
We chose streamlines that seemed to minimize the high density spikes, but were not able to eliminate them.
Fortunately the afflicted regions, which are
restricted to 
disk edges, are tiny and easily masked out.

We smooth $\Sigma$ by first applying the Python function \texttt{scipy.ndimage.interpolation.map\_coordinates}
which sub-samples to a finer non-regular $r$-$\theta$ grid using a local order-1 spline. We then establish $\Sigma$ on a regular $r$-$\theta$ grid with \texttt{scipy.interpolate.griddata}
which uses a local linear interpolation. 
The final step is to normalize the disk mass 
so that the barycenter remains
at the origin. Figure \ref{Fig2} shows our final gas disk,
of mass $M_{\rm disk} = 0.028 M_\odot$. 

To 0th order, the gas velocities are the (interpolated)
velocities of the test particles used to construct the original 22
streamlines. We add a 1st order correction to account
for gas pressure gradients. With a fractional error of $\sqrt{\mu}$,
pressure forces are balanced by Coriolis forces (the flow is geostrophic); the velocity corrections read
\begin{eqnarray}\label{geostrophic_vel}
		v_{r,1} = -\frac{c_{\rm s}^{2}}{2\Omega r }\frac{\partial \ln \Sigma}{\partial\theta} \\
		v_{\theta,1} = \frac{c_{\rm s}^{2}}{2\Omega}\frac{\partial \ln \Sigma}{\partial r} \label{geostrophic_vel_theta}
\end{eqnarray}
and are evaluated by applying Python's \texttt{gradient} function
to $\Sigma$. 
Minor numerical artifacts in the 1st order velocities
at disk edges are smoothed using a combination of 
median and Gaussian filters. The sound speed $c_{\rm s}$
is fixed at 0.37 km/s, appropriate for gas at 40 K.

\begin{figure*}[htp]
\epsscale{1}
\figurenum{3}
\plotone{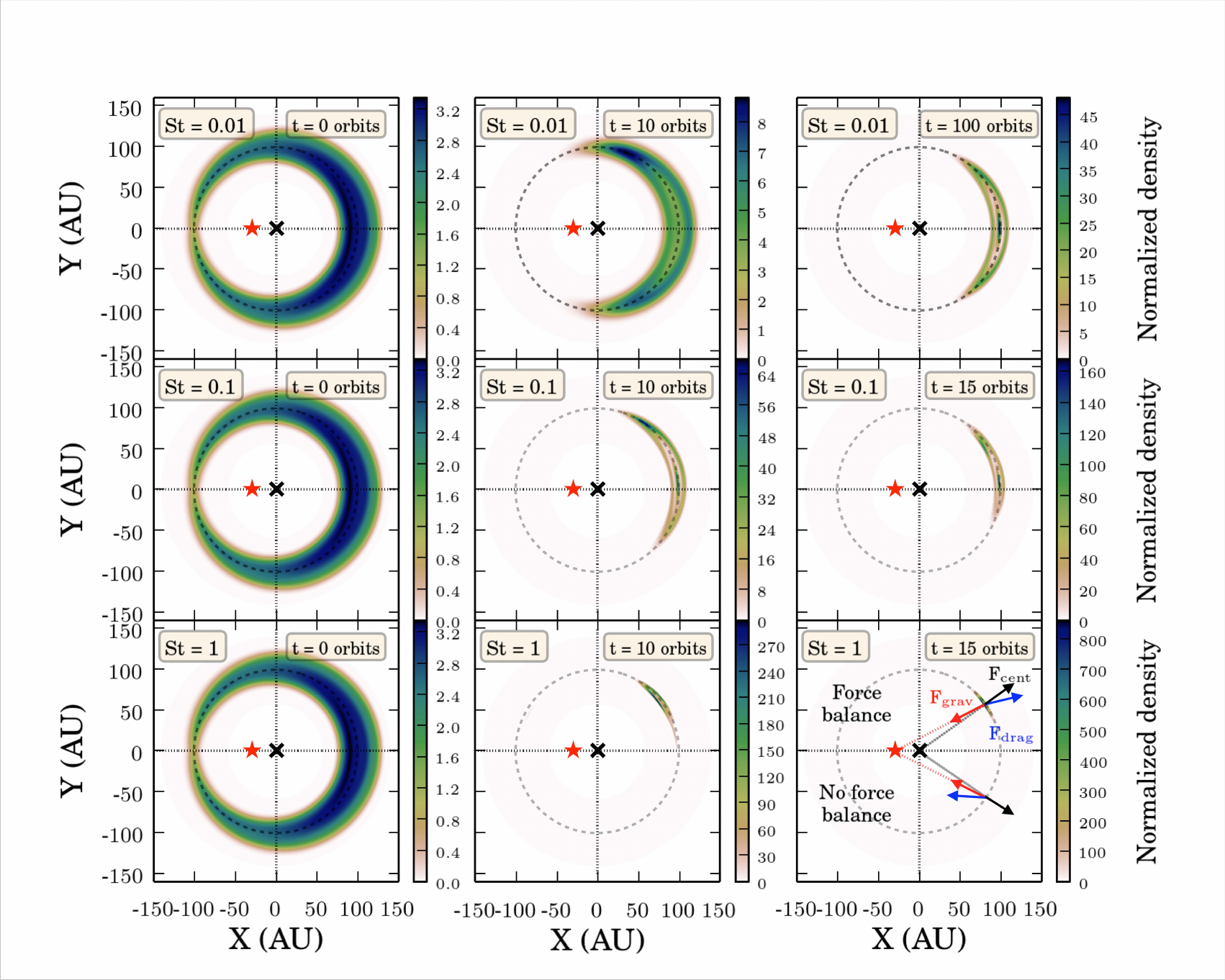}
\label{Fig3}
\caption{How dust surface density varies with time (left to right) and
Stokes parameter (top to bottom). Since our gas disks are not turbulent,
all dust grains of a given size eventually collect to a point.
The location of this point varies with grain size; see also Figure \ref{Fig5}.
For ${\rm St} = 1$, the collection point is displaced from the usual
stable equilibrium by about +45$^\circ$; here
drag (which points radially outward because 
$\partial \Sigma / \partial \theta < 0$ in equation \ref{geostrophic_vel})
can balance gravity and centrifugal forces.
Adding a turbulent diffusivity $\alpha$ can 
keep dust patterns more spatially extended (\S\ref{conc}).}
\end{figure*}

\section{DUST IN GAS}
We now add dust particles to our gas disk. A dust particle
behaves like a test
particle except that it also feels
a gas drag acceleration
$-(\mathbf{v}_{\rm d}-\mathbf{v}_{\rm g})/t_{\rm stop}$
which damps dust-gas relative velocities
over an aerodynamic stopping time $t_{\rm stop}$.
The stopping time scales inversely as the volumetric
gas density $\rho$, which we compute from
$\Sigma$ by assuming the disk has a constant
vertical thickness $h = c_{\rm s}/\Omega$.
Our convention is to assign every particle a Stokes
parameter ${\rm St} \equiv \Omega t_{\rm stop}$ 
evaluated at the position of peak $\rho$ (i.e., at the stable
point $x=x_{\rm eq}$, $y=0$); the stopping time at any other
location then scales inversely as the local $\rho$
(which deviates from the peak value
by at most a factor of $\sim$3).
A total of 2100 dust particles are laid down
with an initial surface density
matching that of the gas disk and with initial velocities
equal to the 0th order gas velocities. 
Dust trajectories are computed using the same Runge-Kutta
integrator as was used to obtain the gas streamlines.
Gas densities and velocities are computed by local
linear interpolation.

\begin{figure}[h]
\epsscale{1.}
\figurenum{4}
\plotone{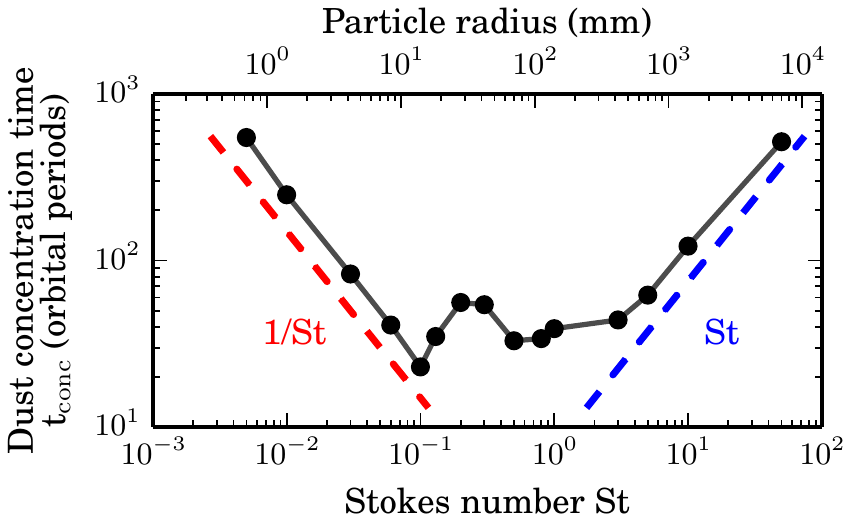}
\label{Fig4}
\caption{Dust concentration timescales (defined as the time for 
dust that is initially well-mixed with gas
to collect into a region subtending 4$^\circ$ in azimuth).
These simulation results confirm the asymptotic scalings predicted by equation
(\ref{time_conc}) for ${\rm St} \ll 1$ and ${\rm St} \gg 1$.}
\end{figure}

Figure \ref{Fig3} shows the evolution of dust surface density
for ${\rm St} = 0.01$, $0.1$, and $1$ (equivalent to grain sizes
of $\sim$ 1 mm, 1 cm, and 10 cm for an assumed bulk density of 1 g/cm$^3$).
Because gas feels pressure forces while dust does not,
dust and gas velocities generally differ.
Dust particles feel tailwinds/headwinds that drag
them toward gas pressure maxima (e.g.,
\citealt{2010AREPS..38..493C}). For our disk, the
global pressure maximum is located at the stable point,
and indeed particles with ${\rm St} = 0.01$ and $0.1$ collect there.
We estimate the concentration time as follows.
In a gas disk with a radial pressure gradient, the radial speed
of a particle is 
\begin{equation}
v_{{\rm dust},r} \sim \frac{c_{\rm s}^2}{\Omega r} \frac{\partial \ln \Sigma}{\partial \ln r} \frac{{\rm St}}{1+{\rm St}^2}
\end{equation}
valid for any ${\rm St}$ (e.g., \citealt{nakagawa1986settling}).
For our problem,
$\partial \ln \Sigma / \partial \ln r \sim \pm 1/\sqrt{16\mu/3}$ for
streamlines to the left/right of the stable point
(see Figure \ref{Fig3}); we have here used the fact
that our disk has a characteristic radial width
$\sqrt{16\mu/3} \,a_0$ (equation \ref{Disk_width_scale}).
Thus particles interior/exterior to the stable point
drift radially outward/inward, traversing a series of ever-smaller
horseshoes. The timescale to cross the radial
width of the disk and collect onto the stable point is then
\begin{equation}\label{time_conc}
t_{\rm conc} \sim \frac{ \sqrt{16\mu/3} \, a_0}{v_{{\rm dust},r}} \sim \frac{16}{3} \frac{\mu}{\Omega} \left( \frac{\Omega a_0}{c_{\rm s}} \right)^2 \frac{1+{\rm St}^2}{{\rm St}} \,.
\end{equation}
which predicts that ${\rm St} \sim 1$ particles
accumulate fastest. Our numerical experiments
confirm this expectation and also verify the
asymptotic scalings ($t_{\rm conc} \propto 1/{\rm St}$ for ${\rm St} \lesssim
0.1$ and $t_{\rm conc} \propto {\rm St}$ for ${\rm St} \gtrsim 4$;
see Figure \ref{Fig4}).

\begin{figure}[h]
\epsscale{1.}
\figurenum{5}
\label{Fig5} 
\plotone{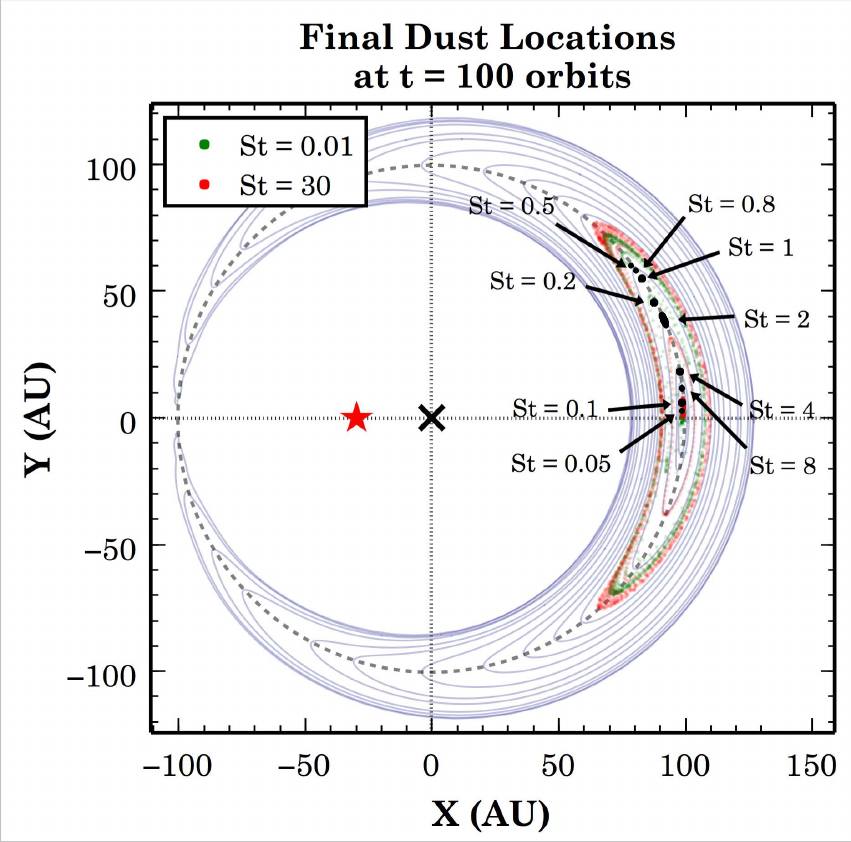}
\caption{Final 
accumulation points for dust particles with various ${\rm St}$
numbers. The collection points advance in azimuth from ${\rm St} = 0.05$ to
0.5, and then move back down from ${\rm St} = 0.5$ to 8. Depending
on the dust size distribution, this systematic variation might be
observable by imaging the disk at multiple wavelengths.
Particles with ${\rm St} = 0.01$ and $30$ have yet to converge to the stable
point at the end of 100 orbits when this snapshot was taken.
}
\end{figure}

Surprisingly, not all particles concentrate at the global
pressure maximum located on the $x$-axis (the gas disk symmetry axis).
Figures \ref{Fig3} and \ref{Fig5} reveal that marginally aerodynamically
coupled particles collect along an arc that extends off
the $x$-axis in the direction of orbital motion. 
Starting from the $x$-axis at ${\rm St} \lesssim 0.1$, 
the position angles of final accumulation points
advance with increasing ${\rm St}$ to a maximum of
$\sim$45$^\circ$ at ${\rm St} \approx 0.5$; further increasing
${\rm St}$ causes the position angles to slide back 
down, returning to zero at ${\rm St} \gtrsim 10$.
The bottom right panel of
Figure \ref{Fig3} illustrates that,
as long as the drag force is neither too strong nor too weak,
a three-way force balance between the offset stellar gravity,
the centrifugal force, and drag is possible, but only in one quadrant
of the rotating frame.
Particle size segregation off-axis offers an observational test
of our model:
the locations of peak intensity in dust emission maps may vary
systematically with wavelength, to the extent that different
observing wavelengths select for different size particles.

\section{DISCUSSION AND FUTURE DIRECTIONS} \label{conc}
A lopsided disk displaces its host star from the system barycenter. In the
resultant ``offset'' stellar potential (one that has an indirect term),
we have calculated that gas streamlines take the form of
horseshoes. Our calculations are not self-consistent
because on the one hand we have invoked the gas disk's mass to move the star,
but on the other we have neglected the disk's gravity when computing gas streamlines. 
Our neglect of gas self-gravity is a severe approximation:
the disk's gravity is as strong as the indirect forcing, since both scale as $\mu$.
Indeed, when we re-calculate {\it a posteriori} the potential $U$ including the gas disk's
contribution, we find that $U$'s topology changes qualitatively.
In particular, what was
a global extremum at $x = x_{\rm eq}$ becomes a saddle point.
Thus the mode we have constructed in this paper is not an equilibrium fast mode.
Still, we might not be too far off the mark; reducing the disk mass by a factor of $\sim$2
(while keeping the stellar offset $\mu$ fixed) restores the original topology at $x_{\rm eq}$. We are also
encouraged by stellar dynamics calculations performed by \citet{2008MNRAS.387....2D}, who
found that their ``rotating lopsided mode" (i.e., a fast $m=1$ mode) supports ``banana"
(i.e., horseshoe-shaped) orbits; see their Figure 14. Future work should incorporate disk
self-gravity, not only to construct self-consistent fast modes,
but also to search for instabilities that can create lopsided mass distributions.

One step toward more realistic gas equilibria is to replace our
test-particle calculation with a hydrodynamical one that accounts
for gas pressure. In this regard, we experimented with a 2D inviscid
hydro-simulation using the \texttt{PLUTO} code 
\citep{2007ApJS..170..228M},
evolving an initially circular and uniform gas ring in an offset
stellar potential (see caption to Figure \ref{Fig2}b for technical
details). A few numerical difficulties 
complicated
this first-cut experiment.
The system did not settle into steady state; gas leaked
outward and was lost off the grid; and our boundary conditions,
chosen mainly for convenience, affected the propagation of spiral density
waves. Nevertheless, $\sim$20--80 orbits into our hydro-simulation,
we observed the gas disk conforming to the same horseshoe shapes
obtained with our test-particle method---compare Figures
\ref{Fig2}a and \ref{Fig2}b---and found consistent results for
dust concentration. The qualitative agreement between our test-particle
calculation and our hydro-simulation helps to validate the former,
and motivates us to improve upon the latter by including gas self-gravity,
viscosity, turbulence, and dust feedback.

Fast modes promise
to better reproduce observed dust distributions with large azimuthal and radial extents
(e.g., \citealt{2013Natur.493..191C}; \citealt{2013Sci...340.1199V}).
In our demonstration model, particles with sizes of 0.1--1 mm have concentration
times on the order of the disk age or longer,
and should thus
occupy horseshoe-shaped regions like those shown in
the top right panel of Figure \ref{Fig3}.
In addition, larger particles, having Stokes parameters
${\rm St} \sim 0.1$--10
(particle sizes of 1--100 cm), naturally spread
themselves in azimuth by $\sim$45$^\circ$.
Gas turbulence can also diffuse dust particles, keeping them
on large horseshoe orbits. The P\'eclet number is the ratio of timescales
for turbulent diffusion and concentration:
\begin{eqnarray}
{\rm Pe} \equiv \frac{t_{\rm diff}}{t_{\rm conc}} \sim \frac{(\sqrt{16\mu/3} \, a_0)^2 / D}{t_{\rm conc}} \sim \frac{{\rm St}}{\alpha} \,.
\end{eqnarray}
Here $D \approx D_{\rm g}/(1 + {\rm St}^2)$
is the dust particle diffusivity
(equation 5 of \citealt{2007Icar..192..588Y}),
$D_{\rm g} \sim \alpha c_{\rm s} h$ is the gas diffusivity,
and $\alpha < 1$ is the Shakura-Sunyaev turbulence parameter.
Small ${\rm Pe} \lesssim 1$ --- i.e., more spatially extended dust --- obtains
for ${\rm St} \lesssim 0.01$ (mm-sized or smaller particles)
and $\alpha \sim 0.01$. 

Our proposed lopsided gas mode is gravitational and global; hence its
radial width is not limited by the pressure scale height $H$. 
This limitation afflicts vortices, since the Keplerian shear across a vortex must remain subsonic
(e.g., \citealt{2009ApJ...693...85L}; \citealt{2014arXiv1405.2790Z}). 
\citet{2013ApJ...775...17L} acknowledge this difficulty
in the last paragraph of their
section 6.5, noting that the radial half-width of the dust trap
in Oph IRS 48 is 17 AU $\approx 3H$ whereas their vortex theory (see their
equation 68) predicts a maximum half-width of $\sim$$H/2$. 
Our horseshoe-shaped orbits are not constrained by $H$ and may thus
better reproduce the azimuthally and radially wide horseshoe-shaped
emission seen in transitional disks like HD 142527
\citep{2013Natur.493..191C}. 
Because the gas mode is global, we speculate
that it may be more resistant to the destructive effects of dust back-reaction than are vortices
\citep{2004A&A...417..361J, 2012MNRAS.422.2399M}. Disk gravity
underlies our mode but may undercut vortices;
\citet{2011MNRAS.415.1445L, 2011MNRAS.415.1426L}
found that self-gravity shrinks vortices and frustrates their merging.

We have not identified the origin of the fast mode.
It may be that, like the vortices of \citet{2014arXiv1405.2790Z},
a planet is responsible. A passing star might also pull the disk to one side of the star
and excite a long-lived fast mode
(see \citealt{2012MNRAS.421.2368J} who use a passing star to excite slow
modes). But for economy of hypothesis, one can do no better than look to the self-gravitational field of the gas disk itself \citep{2008MNRAS.387....2D}.

\acknowledgements
We thank Robin Dong, Mir Abbas Jalali, Eve Lee, Chris Ormel, Roman Rafikov,
and Scott Tremaine for useful discussions, and an anonymous referee for an
encouraging report. EC acknowledges support from the
Miller Institute and a NASA Origins grant.

\end{document}